# Intrinsic interfacial van der Waals monolayers and their effect on the high-temperature superconductor FeSe / SrTiO$_3$

Hunter Sims[1,2], Donovan N. Leonard[2], Axiel Yaël Birenbaum[2], Zhuozhi Ge[3], Tom Berlijn[4], Lian Li[3], Valentino R. Cooper[2], Matthew F. Chisholm[2], and Sokrates T. Pantelides[1,2]

1. Department of Physics and Astronomy, Vanderbilt University, Nashville TN 37235, USA
2. Materials Science and Technology Division, Oak Ridge National Lab, Oak Ridge TN 37831, USA
3. Department of Physics and Astronomy, West Virginia University, Morgantown, WV 26506, USA
4. Center for Nanophase Materials Sciences and Computational Science and Engineering Division, Oak Ridge National Lab, Oak Ridge TN 37831, USA

**Abstract**

The sensitive dependence of monolayer materials on their environment often gives rise to unexpected properties. It was recently demonstrated that monolayer FeSe on a SrTiO$_3$ substrate exhibits a much higher superconducting critical temperature $T_c$ than the bulk material. Here, we examine the interfacial structure of FeSe / SrTiO$_3$ and the effect of an interfacial Ti$_{1+x}$O$_2$ layer on the increased T$_c$ using a combination of scanning transmission electron microscopy and density functional theory. We find Ti$_{1+x}$O$_2$ forms its own quasi-two-dimensional layer, bonding to both the substrate and the FeSe film by van der Waals interactions. The excess Ti in this layer electron-dopes the FeSe monolayer in agreement with experimental observations. Moreover, the interfacial layer introduces symmetry-breaking distortions in the FeSe film that favor a $T_c$ increase. These results suggest that this common substrate may be functionalized to modify the electronic structure of a variety of thin films and monolayers.

In his Nobel lecture, Herbert Kroemer opened with the statement "Often, it may be said that the interface is the device" [1]. Nowhere is this more true than in two-dimensional materials. The band gap of graphene provides an apt example. It is on the order of µeV in the freestanding material [2], arising from spin-orbit coupling, but reaches tens of meV on Cu(111) or hexagonal BN [3] or hundreds of meV in bilayer graphene [4]. Interactions in a 2D material exhibit a strong dependence on the dielectric environment in neighboring substrate or vacuum layers [5]. Perhaps the most surprising example of substrate dependence in a two-dimensional material is the recent discovery of an order-of-magnitude increase in superconducting temperature when monolayerFeSe is grown on SrTiO$_3$ (STO) [6–9]. Similar results have been obtained on BaTiO$_3$ [10] and both anatase and rutile TiO$_2$ [11,12] substrates, but the effect is absent on Bi$_2$Se$_3$ [13] and on graphene [14], where $T_c$ instead decreases as the thickness decreases (as in other superconducting thin films such as Pb [15]). Bulk FeSe, the limiting case of the intercalated iron-pnictide/-chalcogenide system, exhibits a $T_c$ of only about 8 K [16] (reaching 37 K under pressure [17]), In FeSe/SrTiO$_3$ $T_c$ increases by roughly an order of magnitude to 60 – 80 K, with one report reaching above 100 K [18].

We conclude that this enhanced superconductivity is directly related to the interaction between the FeSe monolayer (ML) and its substrate. Theoretical investigations into the role of the substrate have focused on the coupling between electronic states in the FeSe film and phonons in SrTiO$_3$ [19–22] or on charge transfer or doping between the substrate and monolayer [21]. Huang and Hoffman noted, that the structure of the interface between FeSe and STO has not been definitively established [23]. Li *et al.* [24] reported scanning-transmission-electron-microscopy (STEM) Z-contrast images revealing a pair of TiO$_x$ layers, similar to the previously reported double-layered reconstruction of the STO(001) surface [25,26]. Around the same time, Zou *et al.* [27] proposed that the $\sqrt{13} \times \sqrt{13}$ STO(001) surface reconstruction persists in an interfacial TiO$_x$ double layer. However, these proposed double-layer interfacial structures do not fully match the features of the Z-contrast images of Ref. [24]. They do not account for the large spacing that is evident between the proposed pair of TiO$_x$ layers in the STEM images of Ref. [24]. The precise atomic structure of the interfacial layer is necessary for any explanation of the emergent properties of FeSe/STO.

In this Letter, we establish the structure of the interface: a Ti$_{1+x}$O$_2$ interlayer bonded to both the substrate and the FeSe ML by van der Waals (vdW) interactions, accounting for the observed large separation between the interlayer and the substrate. *Both Ti$_{1+x}$O$_2$ and FeSe are essentially two-dimensional monolayers floating above the substrate, making* SrTiO$_3$/Ti$_{1+x}$O$_2$/FeSe *a van der Waals*

*heterostructure*. We show that the interlayer is best described as a (2x2) $Ti_{1.5}O_2$ layer. We demonstrate that the *excess Ti in the interlayer* is responsible for the vanishing of the Fermi surface at the zone center, as observed by angularly resolved photoemission spectroscopy (ARPES) [7,28]. Further, we find that the interlayer breaks the in-plane $C_4$ symmetry of the FeSe film. This lower symmetry stabilizes an in-plane distortion of the FeSe lattice similar to that predicted by Coh *et al.* to enhance $T_c$ [29]. The 50% excess Ti provides a sufficient doping level to fill the $\Gamma$ hole pocket in the FeSe band structure, where the doping level is inversely proportional to the strength of the bonding between the $Ti_{1+x}O_2$ and FeSe layers. Finally, we show that a floating, vdW-bonded $Ti_{1+x}O_2$ monolayer is not unique to $FeSe/SrTiO_3$. A similar $TiO_2$-like layer exists in bulk $Cs_xTi_{2-x/4}O_4$ [30]. Previous work on bronze-phase $VO_2$ grown on $SrTiO_3$ reports an extra titanium oxide layer at the interface [31], which we now show is the same as that between FeSe and STO (Figure S6).

High angle annular dark field (HAADF) imaging the $FeSe/SrTiO_3$ interface was performed at 200kV with a Nion UltraSTEM 200 using an illumination half angle of 30 mrad and an inner detector half angle of 65 mrad. Image simulations were carried out using the same parameters within a multislice model [32] including the quantum excitation of phonons model [33], as implemented in the program μSTEM [34]. All density functional theory calculations were performed within the PBEsol generalized gradient approximation [35] using the Vienna *ab-initio* Simulation Package (VASP) [36]. We use the PAW [37] pseudopotentials of Kresse and Joubert [38] and the vdW corrections of Tkatchenko and Scheffler [39]. To improve accuracy, we included Sr 4*s* and 4*p* and Ti 3*s* and 3*p* semi-core states as valence states. We use the DFT+U method of Liechtenstein *et al.* [40] with U = 3 eV for Ti (typical value for bulk systems, e.g. Ref. [41]) and 1.5 eV for Fe (giving reasonable Fe – Se height). We took J = 0.9 eV in both cases. Although reduced dimensionality can affect these parameters, we emphasize that neither the interface properties nor the presence of the in-plane distortion depends on U. Relaxations were performed on a $\Gamma$-centered 4 x 4 x 1 **k**-point mesh using a plane-wave cut-off of 600 eV. Final calculations on the converged structure used an 8 x 8 x 1 **k** mesh, which was sufficient to converge all reported quantities. Our simulation cell consisted of three layers of $SrTiO_3$ with both faces terminated in $TiO_2$, the $Ti_{1+x}O_2$ interlayer (x = 0.5), a single layer of FeSe, and about 18 Å of vacuum. Terminating the back surface at the SrO layer rather than $TiO_2$ did not alter the properties of the interface. Structural relaxations were constrained to the experimental in-plane STO lattice parameter (*a* = 3.905 Å) to reduce the effect of the limited thickness of the STO slab. See Supplemental Information for further technical details of the synthesis, preparation, and characterization of the samples.

In Figure 1a we present a Z-contrast aberration-corrected STEM image of the interface region captured at an acceleration voltage of 200 kV (see Supplemental Figure 1 for STM characterization of the sample). The double titanium oxide termination is clearly visible, and one can discern faint features between the Ti columns in the second layer. These features become more apparent when viewing the intensity profile along the $Ti_{1+x}O_2$ layer (Figure 1, bottom) and along vertical Ti columns (Supplemental Figure 2). The height of the interfacial $Ti_{1+x}O_2$ layer (IL) above the normal $TiO_2$-terminated substrate is 2.55±0.20 Å, which is itself 1.94±0.24 Å above the SrO layer below. Both of these results are within the ranges reported by Li *et al.* [24]. The FeSe monolayer is 3.25±0.20 Å above the IL. The STO – IL distance represents more than a 30% increase in interlayer spacing compared to 1.95 Å in bulk STO. Comparison of this image and that of Ref. [24] with known reconstructions of the bare STO surface (Supplemental Figures 3 – 5) prompts us to seek a different atomic structure for this interfacial layer.

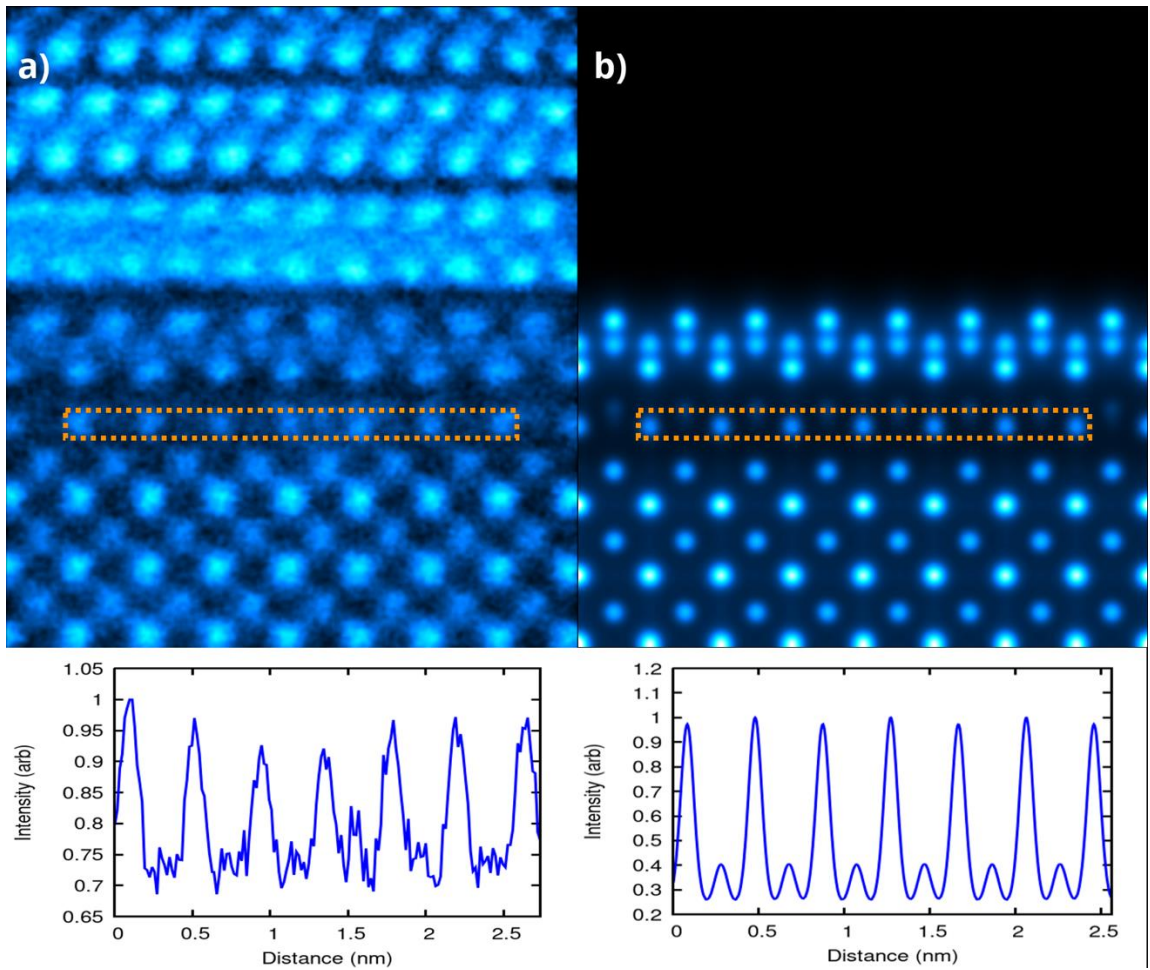

**Figure 1.** (top) Cross-sectional view of the FeSe/SrTiO$_3$ interface. (a) HAADF image, with FeTe capping layers. An additional titanium oxide layer is visible above the standard TiO$_2$-terminated STO surface. (b) Simulated HAADF image using a multislice code [32] based on our (2x2) interface structure (without capping layer). (bottom) Intensity profiles (averaged over a width of 8 pixels or about 13 pm) across the interface layer, showing excellent agreement between the experimental and simulated images of FeSe/Ti$_{1.5}$O$_2$/STO.

Guided by the HAADF images in the present work and those in Ref. [24], we find the following constraints on a structural model for the interface: The alternating dark and bright features require a doubling of the unit cell along at least one direction. The ubiquity of this feature leads to the conclusion that both in-plane directions possess this lowered symmetry, so we require a (2x2) interfacial layer. The increased interlayer distance suggests that the IL must interrupt the expected Ti – O bonding pattern in the STO substrate, i.e., that the cations in the IL are not registered atop the underlying oxygen sublattice (and vice-versa). Consequently, we have constructed a (2x2) IL corresponding to a (½, ½) shift of the normal TiO$_2$-terminated surface with additional Ti ions in half of the square cavities (as depicted in Figure 2a). The full-intensity Ti columns in the IL always sit above the Sr columns, and we enforce this constraint on both the *a* and *b*-axis projections.

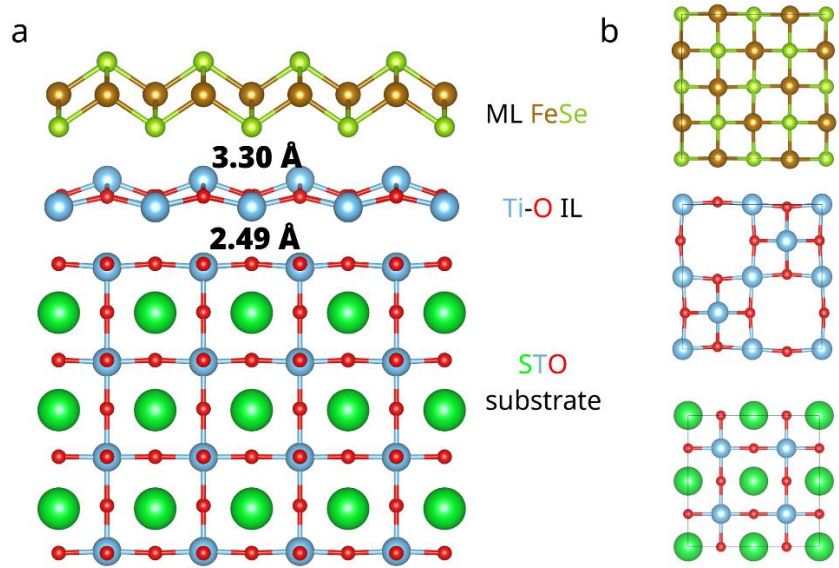

**Figure 2.** Structure of the FeSe/ $Ti_{1.5}O_2$ / $SrTiO_3$ interface, projected along the [100] direction. a) The interface with all atomic positions relaxed. The interlayer distances (2.49 Å and 3.30 Å) compare well with the experimental values (2.55±0.20 and 3.25±0.20 Å) In addition to the increased distance between the substrate and the interfacial layer (2.49 Å compared to the expected 1.95 Å), we find that the terminal $TiO_2$ pulls closer to the layer below (1.82 Å). The extra Ti atoms in the interfacial layer are raised toward the FeSe film. b) A top view of the three components of the interface: FeSe (top), $Ti_{1.5}O_2$ (middle), and $SrTiO_3$ (bottom).

We further refined our model using density functional theory, optimizing the full heterostructure with FeSe initialized in checkerboard antiferromagnetic ordering and all other atomic species unpolarized. We find that it is necessary to introduce van der Waals corrections to prevent the interfacial layer from completely dissociating from the substrate. Applying the Tkatchenko-Scheffler method [39] yields an interlayer distance between the STO substrate and the IL of about 2.49 Å, in excellent agreement with the experimental value of 2.55±0.20 Å (and compared to ~3 Å in the absence of vdW forces). We also find that the distance between the terminal $TiO_2$ layer and the SrO layer below is reduced to 1.82 Å (smaller than but within the error bars of our measurement), in agreement with Ref. [24]. See Figure 2 for a detailed view of the calculated structure. The distance between the Fe and Se sublayers is 1.29 Å, and the Fe-Se-Fe angle is 112.5°, presumably due to the epitaxial strain imposed by the $SrTiO_3$ substrate. The bottom of the FeSe layers sits about 3.3 Å above the IL, again in striking agreement with our measurements (3.25±0.20 Å). To understand how the capping layer might affect the structure, we performed additional calculations with a FeTe layer on top of the FeSe ML (as is the case in the samples we imaged). Depending on the initial separation of the $Ti_{1.5}O_2$, FeSe, and FeTe layers, we found that we could relax structures with IL-FeSe distances of around 2.15 Å as well as 2.49 Å, with some additional distortion in the interface region in the former case. We therefore note that direct comparison of STEM measurements to superconducting properties must be made cautiously. All scanning tunneling microscopy or ARPES experiments are performed *in-situ* with a clean FeSe surface, while the STEM samples have been capped and have undergone further preparation (see the Supplemental Information for details). Further, Rooney *et al.* found that the van der Waals gaps of monolayer selenides are particularly susceptible to discrepancies between experiment and DFT-vdW due to the presence of impurities or undulations at the interface [42]. We therefore rely on experimental images for the atomic structure while studying the effect of variations in interlayer spacing.

A comparison of the HAADF images with our multislice STEM simulations (Figure 1b) shows that the structure of our proposed interface agrees well with the experimental data. When we remove the FeSe monolayer and relax the structure, the $Ti_{1.5}O_2$ returns to the STO surface with the bulk interlayer spacing of 1.95 Å. This agreement leads us to conclude that we have accurately determined the FeSe/STO interface. The presence of the film causes the second titanium oxide layer to lift off from the STO substrate and form a separate two-dimensional interlayer that bonds to both substrate and film via van der Waals interactions. Gao et al. [31] reported a titanium oxide layer between STO and monoclinic bronze-phase $VO_2$ that possesses similar properties (alternating intensities, increased interlayer spacing). In Supplemental Figure 6, we show a relaxed structure using the same methods employed in this study where we find that the IL bonds chemically with the deposited film while still forming vdW bonds with the substrate. Furthermore, layer-resolved electron energy-loss spectra at the interface are consistent with the change in coordination and/or nominal charge state found in our structure (a similar result is obtained by Li *et al.* in FeSe/STO [24]). Given that a floating $Ti_{1+x}O_2$ monolayer also appears between complex oxides with dissimilar symmetries, we hypothesize that such monolayers may provide an alternate path toward epitaxial heterointerfaces that lack continuous perovskite lattice structure.

Our calculations show that the interlayer in the $FeSe/SrTiO_3$ system is not merely a passive glue holding substrate and film together. In fact, the interfacial Ti atoms develop magnetic moments (just under 1 $\mu_B$) and ferromagnetic orientation. Forcing a (necessarily frustrated) antiferromagnetic ordering in the IL yields an excited state about 1.5 meV / Ti higher in energy. Long-range magnetic order at the interface is more likely to reduce $T_c$ and perhaps give rise to vortices, both of which contradict experiment. To explore other possibilities, we performed additional calculations forcing no net spin. Relaxations yielded a similar atomic structure but a reduction of the IL – FeSe interlayer distance to about 2.5 Å. The distance between the IL and the STO substrate is increased to 2.8 Å. This shift in the middle vdW layer accompanies an apparent increase in the electron doping in the FeSe film. Figure 3 illustrates the effect of the IL and of the variation in interlayer spacing on the band structure. We plot the Fe *d*-orbital band structure of a) bare ML FeSe, b) ML FeSe with the $Ti_{1.5}O_2$ IL (with the observed interlayer spacing), c) the full heterostructure with the interlayer spacing observed in STEM, and d) the full relaxed (nonmagnetic) heterostructure. All band structure calculations are computed in the nonmagnetic case, which gives a better agreement with ARPES data [7,28]. We observe a trend in the occupation of the valence band edge at Γ. The IL dopes the FeSe film enough to almost fill the hole pocket, placing the top of the band about 50 meV above the Fermi level (150 meV lower than in ML FeSe). The inclusion of the symmetrically $TiO_2$-terminated STO substrate with the interlayer spacings seen in Figure 2 returns the top of the Γ pocket to about 200 meV above the Fermi level. In contrast, the relaxed structure completely fills the pocket, placing the Γ band about 100 meV below the Fermi level and slightly altering the shape of the M electron pockets. The trend from 3c to 3d can be explained by considering the distances between the IL and ML: a shorter distance reflects stronger bonding and thus a greater capacity for doping. Indeed, it is not clear how meaningful doping could occur across a 3.3 Å vdW bond. The (2x2) periodicity of the $Ti_{1.5}O_2$ supercell can in principle give rise to gap openings in the bands away from the Fermi energy that would be interesting to look for in ARPES experiments and also via the so-called band unfolding theory [43–47]. We do point out that no reconstructions at the Fermi surface are expected given that the **q** vectors associated with this supercell – e.g. $(\frac{\pi}{2a}, \frac{\pi}{2a}, 0), (\frac{\pi}{2a}, 0, \frac{\pi}{2a})$ – are too large to couple states within the electron pockets of the Fermi surface.

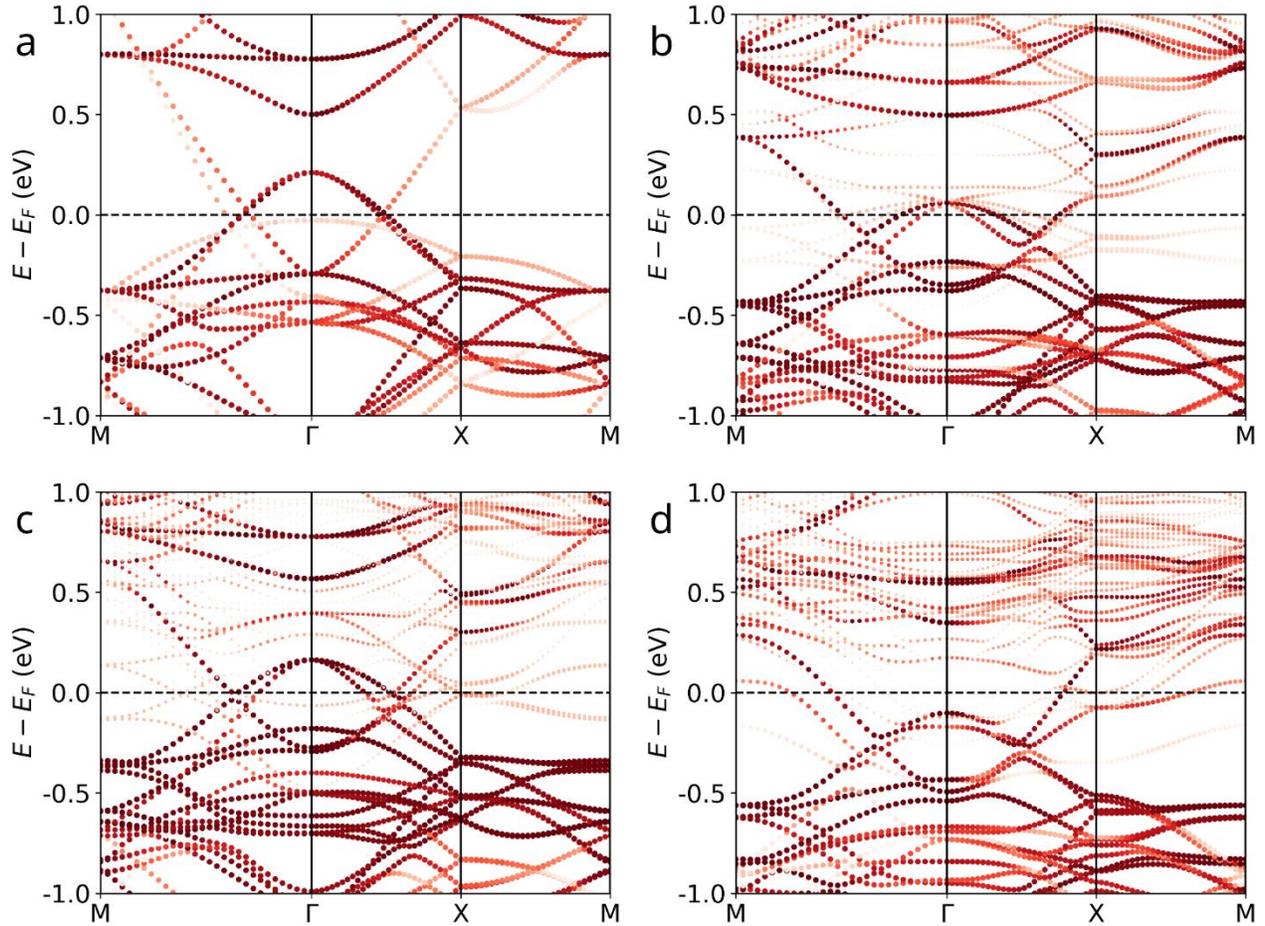

**Figure 3.** Fe *d* band structure of the a) a free-standing FeSe monolayer, b) a FeSe monolayer with a neighboring $Ti_{1.5}O_2$ layer (using the experimental interlayer distance), c) full heterostructure with the interlayer spacings appearing in Fig. 2a, and d) full heterostructure with the calculated interlayer spacings. The addition of the interlayer in (b) nearly fills the hole pocket, leaving only a small Fermi surface around Γ peaking at about 50 meV above the Fermi level. Reintroducing the substrate reverses this trend somewhat (c), but fully relaxing the atomic positions and interlayer distances fully eliminates the hole pocket (d). The amount of Fe *d* orbital character is indicated by the intensity of the red (with the darkest circles corresponding to the largest Fe *d* projection).

In Ref. [48], it was observed that bulk $Fe_{1.01}Se$ undergoes a small orthorhombic distortion below 90 K such that two Fe – Fe distances emerge, differing by about 1.5 pm. We find that similar but larger distortions arise naturally from the true interface structure (as seen in Figure 4). These distortions are absent from the calculated structure when the FeSe monolayer is placed on the typical $TiO_2$-terminated STO substrate (i.e. without the interfacial layer). The $Ti_{1.5}O_2$ interlayer breaks the $C_4$ symmetry of the

FeSe layer, leaving only $C_2$ symmetry along the supercell diagonal. Consequently, we find that the Fe atoms shift from their positions in the square lattice, forming alternating "long" and "short" distances differing in length by ~0.2 Å. Calculating the electron-phonon coupling proved infeasible in our 90-atom simulation cell, but Coh and coworkers [29] recently emphasized the importance of these distortions in enhancing the coupling to certain FeSe phonon modes associated with the M electron pocket. Using a modified semi-local potential, they found a similar shear-like distortion in monolayer FeSe and reported a $T_c$ of 20 – 25 K (compared to ~1 K in the undistorted structure within standard GGA). We conclude that the ability of this naturally forming interlayer to enhance pairing interactions through electron-phonon coupling suggests that it plays a central role in the superconducting properties of FeSe/STO, FeSe/BaTiO$_3$, and FeSe/TiO$_2$. In addition it has been argued that the superconductivity in monolayer FeSe can be enhanced by a coupling between the Fe-d electrons and the phonons in the substrate that is peaked at small momenta [8,22,49]. Our proposed interface will allow for a better understanding of this mechanism.

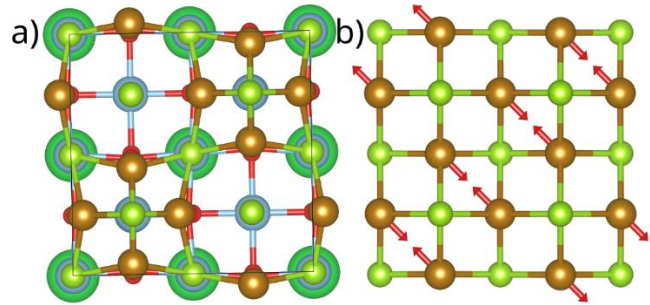

**Figure 4.** View down the [001] direction of the FeSe / Ti$_{1+x}$O$_2$ / SrTiO$_3$ interface. a) Complete relaxed structure with exaggerated Fe distortions. One notes a distortion of the FeSe lattice such that neighboring diagonals are alternatingly closer or farther apart. For the diagonals along [110], the short distance is 2.68 Å and the long distance is 2.82 Å. Along the perpendicular [1$\underline{1}$0] direction, these distances are 2.72 Å and 2.80 Å, respectively. Fe atoms show a tendency to pull closer to the raised Ti sites, with the Fe – Fe distances along [100] and [010] being 3.83 Å above such a site and 3.98 Å otherwise. b) An exaggerated schematic of the distortions between the [110] diagonals.

In summary, we have determined that the naturally occurring double titanium oxide surface reconstruction on STO (001) forms a (2x2) Ti$_{1.5}$O$_2$ layer at the interface between SrTiO$_3$ and monolayer FeSe. This interfacial layer is bonded to both substrate and film by van der Waals forces. Our DFT+vdW calculations show that this layer, which appears and facilitates epitaxy in at least one other complex oxide heterointerface, provides the doping level needed to partially fill or eliminate the Γ hole pocket as observed in ARPES measurements. Further work is needed to determine the interlayer spacing present in the superconducting system. This layer also supports an in-plane distortion in the FeSe ML. This van-der-Waals-bonded interlayer is therefore essential for a full understanding of the superconducting properties of this system and should be included in future theories. Further theoretical and experimental investigation is required – particularly of the phonon properties of this interfacial layer – to fully elucidate the role of this interfacial layer in the electronic and magnetic properties of FeSe and to see whether its effect can be replicated in other layered superconductors.

**Acknowledgements:**
DFT calculations and related analysis were supported by U.S. DOE grant DE-FG02-09ER46554 and the McMinn endowment at Vanderbilt University (HS, STP). They were performed at the National Energy Research Scientific Computing Center, a DOE Office of Science User Facility supported by the Office of Science of the U.S. Department of Energy under Contract No. DE-AC02-05CH11231. Work at ORNL is sponsored by the U.S. Department of Energy, Office of Science, Basic Energy Sciences, Materials Sciences and Engineering Division (DNL, AYB, TB, VRC, MFC). Work at West Virginia University (ZG, LL) is supported by the U.S. National Science Foundation, Division of Materials Research (DMR-1335215). HS wishes to acknowledge helpful communication with G.-X. Zhang and A. Tkatchenko concerning the application of vdW-TS to PBEsol. We also wish to acknowledge useful conversations with T. A. Maier concerning the theory of superconductivity.